\documentclass[superscriptaddress,twocolumn,showpacs,
amssymb,amsmath,nobibnotes,aps,prd,
nofootinbib]{revtex4-1}
\pdfoutput=1
\usepackage{graphicx,subfigure,bm,color,psfrag,hyperref}

\usepackage{amsfonts}
\usepackage{lipsum}
\usepackage{tikz}
\usetikzlibrary{positioning, arrows.meta, shapes, backgrounds}
\usepackage{mathtools}
\usepackage{verbatim}
\usepackage[normalem]{ulem}
\hypersetup{colorlinks,linkcolor={blue},citecolor={red},urlcolor={magenta}}  
\usepackage{lipsum}

\begin{document}

\title{Generic Predictions for Primordial Perturbations and their implications
}
\author{Mohit K. Sharma}
\email{mrmohit254@gmail.com}
\affiliation{Centre For Cosmology and Science Popularization (CCSP),
        SGT University, Gurugram,  Haryana- 122505, India}

\author{M. Sami}
\email{samijamia@gmail.com}
\affiliation{Centre For Cosmology and Science Popularization (CCSP),
        SGT University, Gurugram,  Haryana- 122505, India}
\affiliation{Eurasian International Centre for Theoretical Physics, Astana, Kazakhstan}
\affiliation{Chinese Academy of Sciences,52 Sanlihe Rd, Xicheng District, Beijing}
\author{David F. Mota}
\email{d.f.mota@astro.uio.no}
\affiliation{Institute of Theoretical Astrophysics, University of Oslo, P.O. Box 1029 Blindern, N-0315 Oslo, Norway}

\begin{abstract}
We introduce a novel framework for studying small-scale primordial perturbations and their cosmological implications. The framework uses a deep reinforcement learning to generate scalar power spectrum profiles that are consistent with current observational constraints. The framework is shown to predict the abundance of primordial black holes and the production of secondary induced gravitational waves. We demonstrate that the set up under consideration is capable of generating predictions that are beyond the traditional model-based approaches.
\end{abstract}

\maketitle

\section{Introduction}

In exploring the origins of gravitational waves and the abundance of 
primordial black holes within the standard inflationary paradigm, a pivotal 
question arises: \textit{How can we best realize these phenomena in a 
model-independent manner, and could artificial intelligence (AI) contribute to 
this pursuit?} Since the inflationary paradigm have been successful in 
explaining the origin of density perturbations, flatness of the universe, 
and the horizon size, it also holds the possibility of anticipating 
the production of gravitational waves (GWs) and the dense objects like 
primordial black holes (PBH) 
\cite{Bird:2016dcv,Carr:1974nx,Carr:1975qj,Sasaki:2016jop,Green:2020jor,Garcia-Bellido:2017mdw,Germani:2017bcs,Carr:2021}. 
However, the usual way to achieve this 
involves a lot of careful adjustment in the parameter space, and the 
need for reverse cosmological modeling to match some of the 
observational evidence often makes the whole picture quite rigid and 
less general. In order to alleviate these tight restrictions and 
explore other possibilities within a given framework, we opt for AI based technique.

It may be noted that the AI algorithms have been successful in diverse 
problem-solving scenarios, their application presents a viable approach 
to achieve optimal or near-optimal outcomes for the aforementioned 
inflationary challenges. In order to obtain the outcomes for the standard 
inflationary scenario, which contains several theoretical constraints, 
we opt to develop a model-independent framework capable of 
accommodating and addressing necessary requirements, such as 
Cosmic Microwave Background (CMB) constraints 
\cite{Planck:2020,Planck:2018vyg}. 
One effective approach 
to constructing a framework is leveraging the benefits of the 
dimensionless nature of slow-roll parameters. These parameters not 
only offer insights into the evolution of the background spacetime 
but also play a crucial role in determining the amplitude of 
primordial density fluctuations 
\cite{Mishra:2019pzq,Ozsoy:2023ryl,Choudhury:2023,Bhattacharya:2023ysp,Gangopadhyay:2020bxn}. 

The framework comprises a large $m \times n$ grid structure, representing 
the entire range of magnitudes the first slow-roll parameter can assume. 
Each transition from one cell to another is treated as a transition in 
time, effectively determining the second slow-roll parameter at a specific 
time instance. Being unconstrained from any sort of pre-requirements, 
the framework is capable to make predictions while satisfying the 
observational constraints. In order to make the predictions, we use 
the state-of-the-art reinforcement algorithm: 
Proximal Policy Optimization (PPO) of \texttt{OpenAI} 
\cite{ppo}. 
The main reason to choose this algorithm is its proven track record 
of being successful in tackling complex problems and its smooth 
training process. After constructing the framework and completing 
the training process, the model's predictions can be evaluated to 
determine if they exhibit any enhancement in the scalar power 
spectrum and satisfy the inflationary requirements. 

In this paper, our primary goal is not only to examine the predictions 
of the trained model but also to assess if, at the observational level, 
it reveals evidence that may not be feasible within conventional 
cosmological models. Given that the trained model generates 
predictions without imposing constraints on its predictive nature, 
it is likely to produce results that are more natural. 
This is due to the fact that in the standard setup, 
inflationary scalar field potential parameters are needed to be chosen 
very precisely, often to three or four decimal places, to create large 
fluctuations during inflation 
\cite{Cole:2023wyx,Geller:2022nkr}. 
Additionally, in models designed to produce multiple enhancements in the 
primordial power spectrum, selecting parameter values becomes even more 
difficult. One consequence of this reliance on finely-tuned parameters 
is that models predicting PBHs for a specific 
mass scale may not account for the possibility of PBHs with other masses. 
Therefore, to enhance the predictability of PBH production across different 
mass ranges and without the need of fine-tuning, we adopt a deep 
reinforcement learning framework.
Given its flexible nature, this approach not only enables the 
prediction of the abundance of primordial black holes (PBHs) and 
gravitational waves (GWs) generated from the scalar 
perturbations but also allows to examine their distribution. 

The paper is organized as follows: ($i$) We provide a brief
explanation of the working principles of reinforcement learning, 
($ii$) We present a detailed discussion on cosmological inflation 
within a general framework, and discussing the conditions for 
large density fluctuations, ($iii$) We give a detailed explanation 
of our framework, including the pertinent equations employed to 
analyze various phenomena, ($iv$) We examine the predictive results 
derived from our trained model, and ($v$) We conclude with final 
remarks.

\section{An overview to RL} \label{RL-sec}

The field of learning techniques is primarily categorized into three main types: 
($i$) supervised, ($ii$) unsupervised, and ($iii$) reinforcement learning (RL). 
While supervised and unsupervised learning involve mapping input data to output 
categories and utilizing input data alone, respectively, RL distinguishes itself 
through its unique learning mechanism. RL employs a reward-based technique, 
interacting with the environment and evaluating the quality of state-action pairs 
sequentially. Renowned for its effectiveness in handling highly complex nonlinear 
systems, RL serves as a cornerstone in artificial intelligence. The core 
principle of RL entails a computer agent interacting with the environment, taking 
actions, and receiving feedback. Based on the outcomes of these actions, the 
agent receives rewards or penalties, aiding in the identification of favorable 
actions at each step. The agent's ultimate objective is to maximize the 
accumulated rewards over time.

In a general setup, the agent is guided by an AI algorithm, which, at 
a specific time step $t$, observes a state $s_t$. Executing an action $a_t$ 
results in a transition to state $s_{t+1}$. At each time step, the agent receives 
a reward $r_t$ as feedback. The primary aim of the agent is to acquire knowledge 
of a policy $\pi$ (also denoted as a strategy), facilitating the accumulation of 
the highest possible reward during its exploration in the environment. In each 
iteration of interaction with the environment, the agent updates its policy based 
on the preceding rewards, utilizing this information to discern the optimal 
approach for maximizing overall rewards.

\subsection{Markov Decision Process}

The decision on how does the agent will learn in an environment is described 
by a stochastic process known as the markov decision process (MDP) which 
consist of the following elements:
\begin{itemize}
    \item  Set of states $\mathrm{S}$ and actions $\mathrm{A}$ of the environment.
    \item Transition probability: $\mathrm{T}(s_{t+1}|s_{t},a_{t})$ which maps 
    the state and action at timestep $t$ with the distribution of states at the 
    next time step $t+1$.
    \item Reward function $\mathrm{R(s_t,a_t,s_{t+1})}$ based on the current action $a_t$.
    \item Discount factor: $\gamma \in [0,1]$ which signifies the weightage 
    to be given to the immediate reward.
\end{itemize}
The policy $\pi$ defines a mapping from states to the probability distribution of actions, expressed as $\pi = Pr(a_t|s_t)$. Following each episode, characterized by the completion of a specific number of timesteps, the policy undergoes an update, and the accumulated reward is computed as $\mathrm{R} = \sum_{t=1}^{T} \gamma^t r_t$. The primary objective of this stochastic process is to identify an optimal policy, denoted as $\pi^*$, determined as:
\begin{equation} \label{policy}
\pi^* = \underset{\pi}{argmax} \left(\mathrm{E[\mathrm{R(\pi})]} \right)
\end{equation}
which aims to maximize the expected reward for the agent.

\subsection{Policy Gradient Technique}
There are two ways through which an 
agent can learn: ($i$) by learning from the stored past experiences which is 
known as the offline policy, ($ii$) by directly learning from whatever the agent 
encounters in the environment. In this paper, we will follow the second approach, which is more robust. Once a batch of experience has been used to do a 
gradient update the experience is then discarded and the policy gets improved. Due to this 
reason the second approach is typically more efficient because it only use the collected experience once for doing an update.

The policy on which the current 
agent is choosing gets optimized by directly measuring the gradient of 
the policy with respect to the parameters of the current policy. 
The objective function is defined as follows:
\begin{eqnarray} \label{ppo}
&& L(\theta) = \mathbb{E}\left[\min\left(X \cdot A, \text{clip}\left(X, 1 - \epsilon, 1 + \epsilon\right) \cdot A \right)\right] \,, \\
&& \mbox{such that} \quad X = \frac{\pi_\theta(a_t | s_t)}{\pi_{\theta_\text{old}}(a_t | s_t)} \nonumber
\end{eqnarray}
$A$ is the advantage function, and 
$\theta$ are parameters of the policy that we want to optimize, in fact 
these parameters decide the probability of an 
action in a given state. The quantity $\frac{\pi_\theta(a_t | s_t)}{\pi_{\theta_\text{old}}(a_t | s_t)}$ is the ratio of probability of taking an action in 
a new policy with respect to the old policy, it signifies how much the 
policy has been updated. $A$ is the advantage function which signifies 
how much better or worse the agent's actions when compared with the 
expected return. The clip function ensures that the policy update is 
within a certain range: $[1-\epsilon, 1+ \epsilon]$, where $\epsilon$ 
is a hyperparameter ranges from $0.1$ to $0.2$. Then finally either 
clipped or unclipped function is considered whichever is the minimum. 
The `min' function tries to avoid any large change in the policy update.

\section{A General Setup for Cosmological Inflation}

In this section, we will give a brief overview of the inflationary cosmology 
without resorting to any sort of cosmological model. The duration of 
inflation is characterized in terms of the number of e-foldings $N$ which is generally formulated in terms of the scale factor $a(t)$, 
or the Hubble parameter $H(t)$ (where $t$:= cosmic time) as 
\cite{Mishra:2019pzq,Baumann:2009ds}
\begin{equation} \label{efolds}
    N := \int_{a_{ini}}^{a_{end}} \frac{d a(t)}{a(t)} = \int_{t_{ini}}^{t_{end}} H(t) dt \,.
\end{equation}
Here quantities with subscripts `ini' and `end' refer to the time when inflation 
started and ended, respectively. Typically, the duration of inflation should be 
 $\sim 60$ e-folds to avoid the flatness and horizon problems. The Hubble slow-roll 
parameters during inflation can be expressed as
\begin{eqnarray} \label{slowroll}
    &&\epsilon_{_H} \equiv -\frac{\dot{H}}{H^2} = -\frac{d \ln H}{d N} \,, 
    \qquad \mbox{where} \quad \cdot \equiv \frac{d}{dt}\\
    &&\eta_{_H} = \epsilon_{_H} - \frac{1}{2 \epsilon_{_H}} \frac{d \epsilon_{_H}}{d N} \,,
\end{eqnarray}
such that the slow-roll conditions during inflation are satisfied when 
$\epsilon_{_H}, \eta_{_H} \ll 1$. If $\epsilon_{_H} \simeq 1$, the period of inflation 
gets over. The observational quantities, such as the scalar spectral index 
$n_s$ and tensor-to-scalar ratio $r_{_T}$ can be directly formulated in terms 
of the slow-roll parameters as
\begin{eqnarray}
    &&n_s = 1 - 4 \epsilon_{_H} + 2 \eta_{_H} \,, \\ 
    &&r_{_T} = 16 \epsilon_{_H} \,.
\end{eqnarray}
The constraints on $n_s$ and $r_{_T}$ by Planck TT,TE,EE+lowE+lens+BK15+BAO 
are found to be: $n_s = 0.9668 \pm 0.0037$ and $r_{_T} < 0.058$ at the CMB pivot 
scale $k_\ast = 0.002 \, \text{Mpc}^{-1}$. The scale-dependence of the 
Power Spectrum of the generated curvature perturbations $\mathcal{R}$ during inflation can be expressed as: $\mathcal{P}(k) = \mathcal{A}_s(k/k_\ast)^{n_s-1}$ 
where $\mathcal{A}_s = 2.1 \times 10^{-9}$ is the amplitude of large-scale fluctuations at the pivot scale 
\cite{Planck:2020,Planck:2018vyg}.
\subsection{Large Amplitude Fluctuations}

The large scale observations such as CMB provide us the information of 
very small phase of inflationary paradigm which corresponds to the 
comoving scale that exit at the beginning of the inflation. As a consequence, 
a major portion of the inflationary scenario is still left unprobed 
which may contain interesting physics. 
In particular, this unprobed large phase of inflation can help us 
to explain the origins of some of the constituents of the universe such as 
Dark Matter and also the production of the relic gravity waves, etc.

Despite of the fact that the large-scale universe and inhomogenities 
are consistent with the Planck's $\Lambda$CDM predictions, the small-scale 
large overdensities at some comoving wavenumber $k \gg k_\ast = k_{_{CMB}}$ 
could arise from large density 
fluctuations during inflation 
\cite{Harada:2013,Gangopadhyay:2021kmf,Kimura:2021,Di:2017ndc}. 
Upon exiting the horizon, these fluctuations 
gets freeze out and once they enter inside the horizon, they gets collapse. 
If the density contrast $\delta \equiv \delta \rho / \rho \simeq 10^{-1}$, the 
large density perturbations, after entering inside the horizon in radiation or 
matter era, will gets collapsed to form a PBH. The mass of primordial black 
holes (PBHs) can be determined from the e-foldings at which fluctuations exit 
the horizon. Additionally, if the distribution profile of density perturbations 
is known the fraction of PBHs in the total energy density of the universe 
can be identified. 

Having the potential in explaining atleast the substantial part of DM in 
the form of PBH, if not all, the occurrence of large fluctuations and hence 
the enhanced Power spectrum upto a few orders of magnitude is actually 
needed 
\cite{Geller:2022,DeLuca:2020agl,Firouzjahi:2023xke}. 
The large amplitude of fluctuations in the 
inflation may generate by having ultra slow-roll (USR) inflation for a 
very short period of time (roughly for about a few efolds). However, 
this compulsorily incorporation of a phase or phases of USR, despite 
being able to serve the purpose to explain the DM in the form of PBH, 
and the generation of Secondary Induced GWs (SIGW), requires an immense 
level of fine-tuning, due to which the whole scenario of generating large 
fluctuations seems unnatural. In order to avoid any large fine-tuning(s), 
at least with respect to the choosing the scale of the perturbations for 
large scale fluctuations, one can look for other methods or processes. In 
particular, those methods  which evolves dynamically and search for the 
best-possible and more efficient way to fulfil the task. 

\section{Establishment of the framework}

\begin{figure}
    \centering
    \includegraphics[height= 6cm, width=8.5cm]{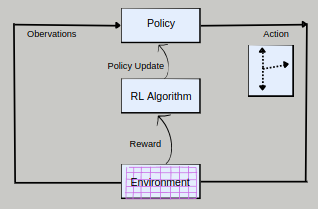}
    \caption{This figure illustrates the Reinforcement Learning technique flowchart.}
    \label{fig:RL}
\end{figure}

In general, the inflationary framework and the one which is 
suited for RL to be implemented are widely disconnected with each other. 
This is due to the fact that the inflation is supposed to be governed by 
a scalar field which has a well-defined potential and kinetic energy, 
whereas, RL works well if there is a system where controlled optimization 
or decision-making is needed. The fact that the general setup of the 
inflation essentially demands the well-defined dynamics 
of the scalar field, aligning it with the required setup for RL poses a 
bigger problem. Not only this, feeding a parameterized form to the 
RL setup will also include the biases and cosmological assumptions 
associated with it, which is undesirable. So, in order to realize the 
inflation as well as to see the possibility of having the enhancement in 
the power spectrum, it becomes necessary to dynamical evolve the system in a 
non-parametric way. This will also ensure that the overall description of 
the system is not fixed beforehand and there can be many possibilities within 
the system.

Since, the whole dynamics of the inflation at every instant can be 
studied with the help of the slow-roll parameters 
(defined in (\ref{slowroll})), they suit appropriate to get utilized 
on the basis of which the overall dynamics can be studied. Moreover, 
they can set to be free from any cosmological model-based assumptions. 
Since, the second slow-roll parameter $\eta_{_{H}}$ can be worked out 
from $\epsilon_{_{H}}$, we only need a single dynamical variable. 
The overall problem now can be setup in a $2-$dimensional Grid-like 
structure which contains a wide range of discrete values of 
$\epsilon_{_{H}}$, approximately $\sim 10^3$. The values are generated 
assuming a uniform distribution given by
\begin{equation}
    \epsilon_{_{H}} \in \mathcal{U}(10^{-10}, 1) \,.
\end{equation}
This range also include possible values required for the enhancement 
of the power spectrum. The problem can now be formulated in the following way:
\begin{itemize}
    \item We first form an environment in the form of a Grid (with 
    dimensions $M \times N$) of 
    $\epsilon_{_{H}}$ values on which the agent will explore many times 
    till it find the best-possible trajectory or trajectories, 
    which is desirable according to our task \footnote{Here, 
    we note that the exploration of the agent extends beyond the neighboring 
    states of the current grid cell. Considering the large transitions 
    needed to achieve large $\eta_{_H}$, we have enabled the agent to take steps 
    of varying sizes. These steps can also alter the state value by an 
    order of magnitude. For more details see Appendix-A{\ref{app1}}.}. 
    \item Each cell in Grid holds a unique value of $\epsilon_{_{H}}$, 
    known as a `state'. In each state, the agent takes the feedback 
    from the environment.
    \item We then define a set of actions
    which takes the agent from one-cell to any other. Each time-step is 
    a measure of e-foldings. Depending on which action the agent will 
    take, the rate of change of $\epsilon_{_{H}}$ will be determined 
    at that particular time-step.
    \item Based on the policy updating rule (\ref{ppo}), the agent will 
    decide itself which action has more probability at a given time-step to 
     enhance the reward $\mathrm{R}$.
    \item The reward is given at each step, specifically if it satisfies the 
    CMB constraints, slow-roll, ultra slow-roll(s), as well as at the last-step 
    if it meets the necessary condition require to end the inflation. 
    \item The policy gets updated after several epochs (generally ranges 
    between $10-100$ depending on the complexity of the system).
    \item The best-policy is then used to predict the outcomes, this works 
    even if the environment gets sightly perturbed.
\end{itemize}
\subsection{Training and Predictions}

\begin{figure}[t!]
    \centering
    \includegraphics[height=6cm, width=8.5cm]{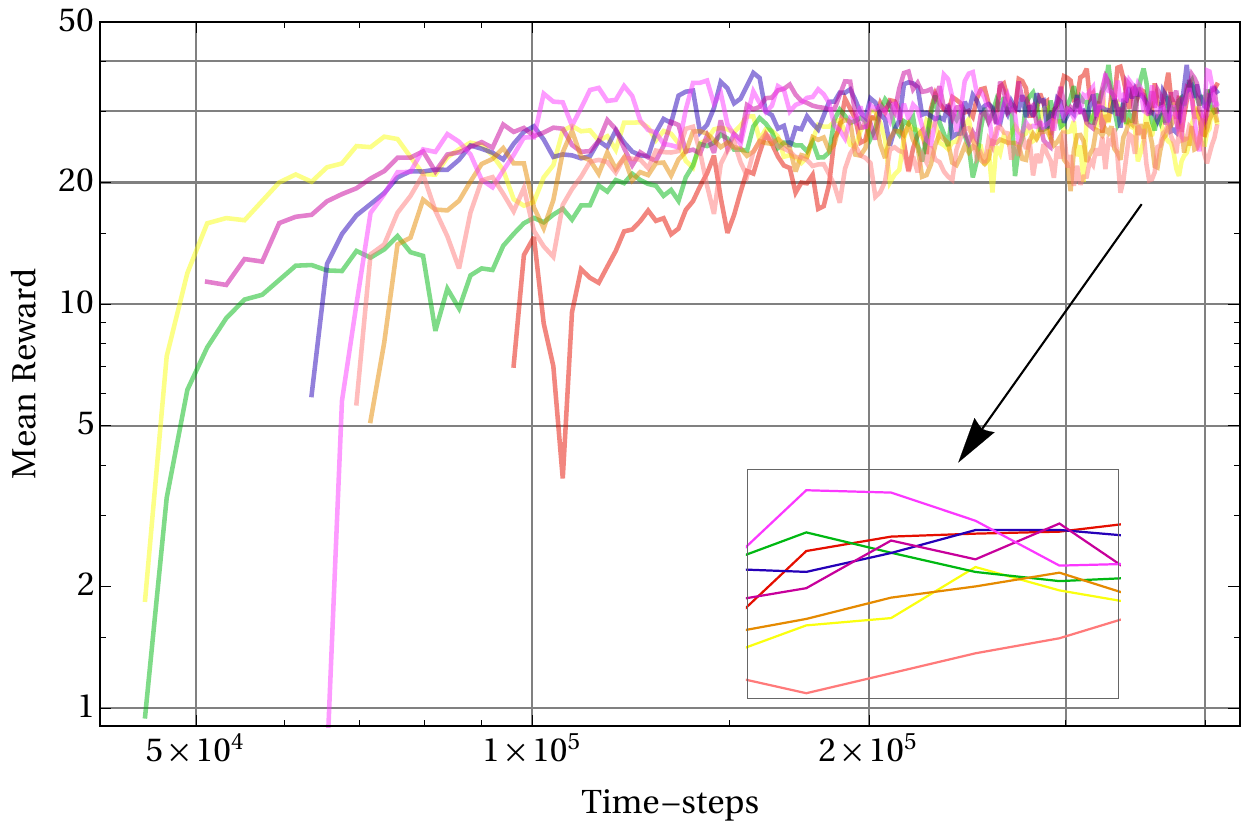}
    \caption{This figure shows the enhancement in the mean reward 
    throughout the training period, encompassing time-steps up to 
    $4 \times 10^5$. Despite originating from various initial reward 
    values, all training sets converge towards a common mean reward 
    as the number of time-steps progresses.}
    \label{fig:reward}
\end{figure}

We train the build model using above-mentioned policy gradient technique 
of \texttt{OpenAI}. One of the reasons to choose this algorithm 
is its reliable performance during training in various domains and scenarios. 
We train our model for roughly $10^5$ time steps. This procedure 
we have followed for a number of times to check if the seed value has 
any impact on the final results. Moreover, we also incorporate 
the randomness in the actions, which takes the agent from one cell to another, 
so that the system can be more further relaxed from any bias. In fig.\,(\ref{fig:reward}) we have plotted 
the mean reward function with the time steps for each training process. 
Here one can see that even if each of the training starts with different 
rewards, they ultimately gets saturated to a particular value. This 
highlights the significance of using this algorithm as a choice and also 
shows that the framework is overall well-defined to make the training 
process smooth. Here, we emphasize that since the environment contains 
rewards and penalties, the agent after getting trained learns about the 
environment and then behaves accordingly. The high reward at the end of the 
training ensures that the agent understands how to progress in such a 
way which gives rise to cosmological inflation in a natural and 
model-independent way i.e. without incorporating any cosmological model
parameters, see Appendix-B\,(\ref{app1}). 

With the trained model, we checked its predictions several times, and 
in each prediction the agent starts with different initial value. 
The trained model, then gives the predictions of slow-roll parameters 
at each small time step, starting from the inflation to its end. 
Since, there is intrinsic randomness involved in the system, 
both in the choice of actions, as well as in the transition between 
the cells for a given action, the system behaves quite differently 
in each prediction while still
satisfying all the necessary requirements. These different behaviours 
open the door for several possibilities which are not realized in 
the conventional way to generate the similar phenomenon, such as 
the formation of PBH and generation of SIGW. 
In order to check whether the predictions fulfil the demands for 
generating the PBH, we first need to solve the famous Mukhanov-Sasaki (MS) equation.

\subsection{Mukhanov-Sasaki Equation}

The MS equation contains the information of the evolution of the 
primordial density fluctuations of each mode. It characterizes the evolution of 
scalar metric perturbations during the inflationary epoch within a 
flat Friedmann-Robertson-Walker (FRW) background. The MS equation 
for the evolution of the comoving curvature perturbation $\mathcal{R}_k$ 
or rather $v\equiv z \mathcal{R}$ ($z$:= MS variable) 
for a given mode $k$ is given by 
\cite{Ragavendra:2020vud,Inomata:2021}
\begin{equation} \label{v-equation}
    v_k''(\tau) + \left[ k^2 - \frac{z''(\tau)}{z(\tau)}\right]v_k(\tau) = 0 \,,
\end{equation}
where $\tau:= \int dt/a(t)$ is the conformal time and $'$ is the derivative 
with respect to $\tau$. By solving the above equation, the corresponding 
dimensionless Power spectrum $P_k$ can be estimated from the following 
relation:
\begin{equation}
    P_k = \frac{k^2}{2 \pi^2}\left|\frac{v_k}{z}\right|^2 \,, 
    \quad \mbox{when} \quad k \ll a(t)H(t) \,.
\end{equation}
For each mode $k$, 
when it is inside the sub-Hubble regime i.e. if $k \gg a(t)H(t)$, one can 
assume the Bunch-Davis condition in which $v$ evolves as
\begin{equation} \label{u-equation1}
    v_k (\tau) = \frac{u_k(\tau)}{\sqrt{2 k}} \exp(-i k \tau) \,,
\end{equation}
where $u_k(\tau)$ is a dimensionless variable. By using (\ref{v-equation}) and (\ref{u-equation1}), one can write the 
second-order differential equation for $u_k(\tau)$ as follows:
$F_k\equiv v_k^2 k+\omega^2_k v^2_k$
\begin{equation} \label{u-equation2}
    u_k''(\tau) - (2i k) u_k'(\tau) - \frac{z''(\tau)}{z(\tau)} \,u_k(\tau) = 0 \,, \\ 
\end{equation}
where
\begin{equation}
\frac{z''}{z} =
    a^2 H^2 \Big[2 - \epsilon_1 + 
    \frac{3}{2} \epsilon_2 + \frac{1}{4}\epsilon_2 ^2
    -\frac{1}{2}\epsilon_1 \epsilon_2 + 
    \frac{1}{2} \epsilon_2 \epsilon_3 \Big] \,,
\end{equation}
and $\epsilon_1 = \epsilon_{_H}$, $\epsilon_{n+1} = -d 
\ln \epsilon_n / dN$ \cite{Bhatt:2022mmn}. 
The above equation (\ref{u-equation2}) can be numerically solved for each 
mode $k$, assuming the initial conditions $u_k(\tau)|_{ini} = 1$ and $u_k'(\tau)|_{ini}=0$. 
Based on the ML predictions on the evolution of slow-roll parameters, 
the resulting power-spectrum profile with $k$ can be obtained for each case.

\subsection{Abundance of Primordial Black Holes}

Let us now focus on the basic criteria for the formation of the PBH 
and their abundance in the universe. As we know that when universe 
enters in the radiation era after the end of the inflation, the fluctuations 
which were generated earlier and gets freeze outside the sub-Hubble region, will 
now enter in later times, depending on when they exited during inflation. 
If the size of these perturbations are very large, they produce large 
curvature perturbation upon re-entering the horizon which may also 
result in the production of the PBH
\cite{Rigopoulos:2021,Papanikolaou:2022}. 
For each mode $k$, PBH of a 
particular mass (in terms of the solar mass $M_\odot$) will be generated 
which is given by 
\cite{Nakama:2016gzw}
\begin{eqnarray}
   && M_{_{PBH}} := \bar{\gamma} M_H = \bar{\gamma}\frac{4 \pi}{3} \frac{\rho_{_{PBH}}}{H^3} \, \\
    && = 1.13 \times 10^{15} \left[\frac{g_\ast}{106.75}\right]^{-1/6} \left[\frac{\bar{\gamma}}{0.2}\right] \left[\frac{k_{_{PBH}}}{k_\ast}\right] ^{-2}
    M_\odot \,, \nonumber
\end{eqnarray}
where $M_H$ is the Hubble mass, $g_\ast$ is the total effective degree 
of freedom of the universe, and $\bar{\gamma}$ is known as the efficiency factor. 

In order to calculate what is the total fraction of the PBHs in the universe 
in the form of the dark matter, we define the following quantity:
\begin{eqnarray} \label{fpbh_1}
    f_{_{PBH}}(M_{_{PBH}}) := \frac{\rho_{_{PBH}}}{\rho_{DM}} = 
    \beta (M_{_{PBH}}) \frac{\rho_{_{T}}}{\rho_{_{DM}}} \,,  \\ \quad \mbox{where} 
    \quad \beta (M_{_{PBH}}) := \frac{\rho_{_{PBH}}}{\rho_{_T}}\Big|_{formation} \nonumber
\end{eqnarray}
and $\rho_{_T}=3H^2 M_{pl}^2$ is the total energy density of the universe. 
The Eq.\,(\ref{fpbh_1}) can also be expressed as:
\begin{eqnarray} \label{fpbh}
    f_{_{PBH}}(M_{_{PBH}}) &&= 1.68 \times 10^8 \left(\frac{\bar{\gamma}}{0.2}\right)^{1/2} \left[\frac{g_\ast}{106.75}\right]^{-1/4} \,, 
    \nonumber \\
  && \times  \left[\frac{M_{_{PBH}}}{M_\odot} \right]^{-1/2} \beta (M_{_{PBH}}) \,. 
\end{eqnarray}
Here we can see from the above equation that the fraction of PBH at present epoch $t=t_0$ as a 
candidate for DM linearly depends on the mass fraction 
$\beta (M_{_{PBH}})$. In order to calculate the latter, we resort ourselves 
to the Press-Schechter formalism which defines $\beta (M_{_{PBH}})$ as the 
probability that the normally distributed density perturbations are above 
a particular threshold value $\delta_c$
\cite{Young:2013oia,DeLuca:2022rfz,DeLuca:2023}
\begin{equation}
    \beta (M_{_{PBH}}) = \bar{\gamma} \int_{\delta_c}^1 
    \frac{d \delta}{\sqrt{2 \pi}\sigma^2} \exp\left(-\frac{\delta^2}{2 \sigma^2}\right) \,,
\end{equation}
where the standard deviation of the coarsed-grained density contrast 
$\delta$ is denoted
by $\sigma$, which is expressed as 
\cite{Franciolini:2018vbk}
\begin{eqnarray}
    \sigma^2 \!&=&\! \frac{16}{81}\int_0^\infty  d \ln k \left( \frac{k}{k_{_{PBH}}} \right)^4 \left[W\left( \frac{k}{k_{_{PBH}}} \right)\right]^2 P_k  \quad \\
   && \mbox{such that} \quad W(x):= \exp\left(\frac{-x^2}{2} \right) \,. \nonumber
\end{eqnarray}

\subsection{Induced Gravitational Waves by Scalar Perturbations}

The gravitational wave (GW) spectrum is characterized by the energy 
density of gravitational waves $(\rho_{_{GW}})$ per logarithmic unit of wavenumber, 
and it is normalized by the total energy density of the universe 
\cite{Kawasaki:2013xsa,Gorji:2023,Vagnozzi:2023}
\begin{eqnarray}
    \Omega_{_{GW}}(k, \tau) &=& \frac{1}{\rho_{_{T}}(\tau)} 
    \frac{d \rho_{_{GW}}}{d \ln k} \,,  \nonumber \\
    &=&   
    \frac{1}{24} \left( \frac{k}{a(\tau) H(\tau)}\right)^2 \mathcal{P}_h(\tau, k) \,,
\end{eqnarray}
where $\mathcal{P}_h(\tau, k)$ is the dimensionless power 
spectrum of the tensor perturbations which in terms of the 
Fourier mode of the transverse-traceless part of the metric 
perturbations $h_{ij}$ can be expressed as
\begin{equation}
   \left(\frac{k^3}{2 \pi^2}\right) \left[<\!\!h^\alpha_\textbf{k} (\tau) 
   h^\beta_{\Bar{\textbf{k}}} (\tau) \!\!> \right]
    = \delta_{\alpha\beta} \, \delta^3(\textbf{k}+\Bar{\textbf{k}}) \mathcal{P}_h(\tau, k)  \,,
\end{equation}
with $\alpha$ and $\beta$ being the two polarization states. 
Following the refs. 
\cite{Bhattacharya:2020lhc,Papanikolaou:2022did}
%
the induced tensor power spectrum $\mathcal{P}_h(\tau, k)$ from the 
second-order scalar density perturbations can be written as 
\begin{eqnarray}
    \mathcal{P}_h(\tau, k) &=& 4 \int_0^\infty da 
    \int_{|1-a|}^{1+a} db \left(\frac{4 a^2 - (1+a^2 -b^2)^2}{4ab} \right)^2 \nonumber \\ \hfill &\times& \mathcal{T}^2(a,b,x) P_\zeta(k a)P_\zeta(k b) \,.
\end{eqnarray}
Here $x = k \tau$, and $\mathcal{T}^2(a,b,x)$ is the integral kernel of secondary induced GWs 
\cite{Baumann:2007}. 
To observe the $\mathcal{P}_h(\tau, k)$ at present 
(when $x \to \infty$) for the modes which entered during the radiation-dominated era ($w=1/3$), the oscillating average of the integral 
kernel can be expressed as 
\cite{Franciolini:2023pbf}
\begin{eqnarray}
   && \bar{\mathcal{T}^2(a,b,\infty)} = \frac{1}{2} 
    \left( \frac{3(a^2 + b^2 - 3)}{4 (ab)^3x}\right)^2 \, \nonumber \\
  && \times \Bigg\{ \left[(a^2 + b^2 - 3)\log\left| \frac{3 - (a+b)^2}{3- (a-b)^2} \right| - 4 ab \right]^2 \nonumber \\ 
     &&+ ~ \pi^2 (a^2 + b^2 - 3)^2 \theta (a+b -\sqrt{3}) \Bigg\} \,,
\end{eqnarray}
where $\theta$ is the Heaviside theta function. The spectrum of the 
gauge-independent GWs that is observed today $a(t)=1$ and produced 
in the radiation era, is given by 
\cite{Balaji:2023ehk,Maiti:2024nhv}
\begin{equation} \label{GW_spectrum}
    \Omega_{_{GW}}^{(0)} h^2 = \Omega_{r}^{(0)}h^2
    \left( \frac{g_{\ast}(T_{rh})}{106.75}\right) \left( \frac{g_{\ast s}(T_{rh})}{106.75}\right)^{-4/3} \Omega_{_{GW}} (k,\tau_r) \,.
\end{equation}
The quantities $g_{\ast}$ and $g_{\ast s}$ represent the effective 
degrees of freedom in the energy density and entropy evaluated 
at the reheating epoch, and $\tau_r$ represent the conformal time 
when a mode enters inside the horizon in radiation era.

\section{Predictive Outcomes}

The predictions generated by the trained RL model offer an opportunity to 
assess its significance in relation to the observational phenomena discussed 
in the previous section. In contrast to binary predictions, denoted as 'yes' 
or 'no', our model is specifically trained to forecast the temporal profiles 
of inflationary slow-roll parameters. These outcomes not only exhibit 
variations in their evolutionary patterns with each prediction but can 
also lead to entirely different scenarios or introduce features that may 
be absent in other predictions. Consequently, one anticipates a diverse 
range of evolutionary slow-roll outcomes, characterized not only by their 
novelty but also by their adherence to the standard inflationary constraints 
suggested by observational data. Moreover, this approach provides us with a 
statistical measure, enabling the calculation of the probability distribution 
function. This capability gives the method an advantage in predicting the 
distribution function of features of inflationary paradigm.

\subsection{Outcomes for Scalar Power Spectrum}

\begin{figure}
    \centering
    \includegraphics[height=6cm, width=8.5cm]{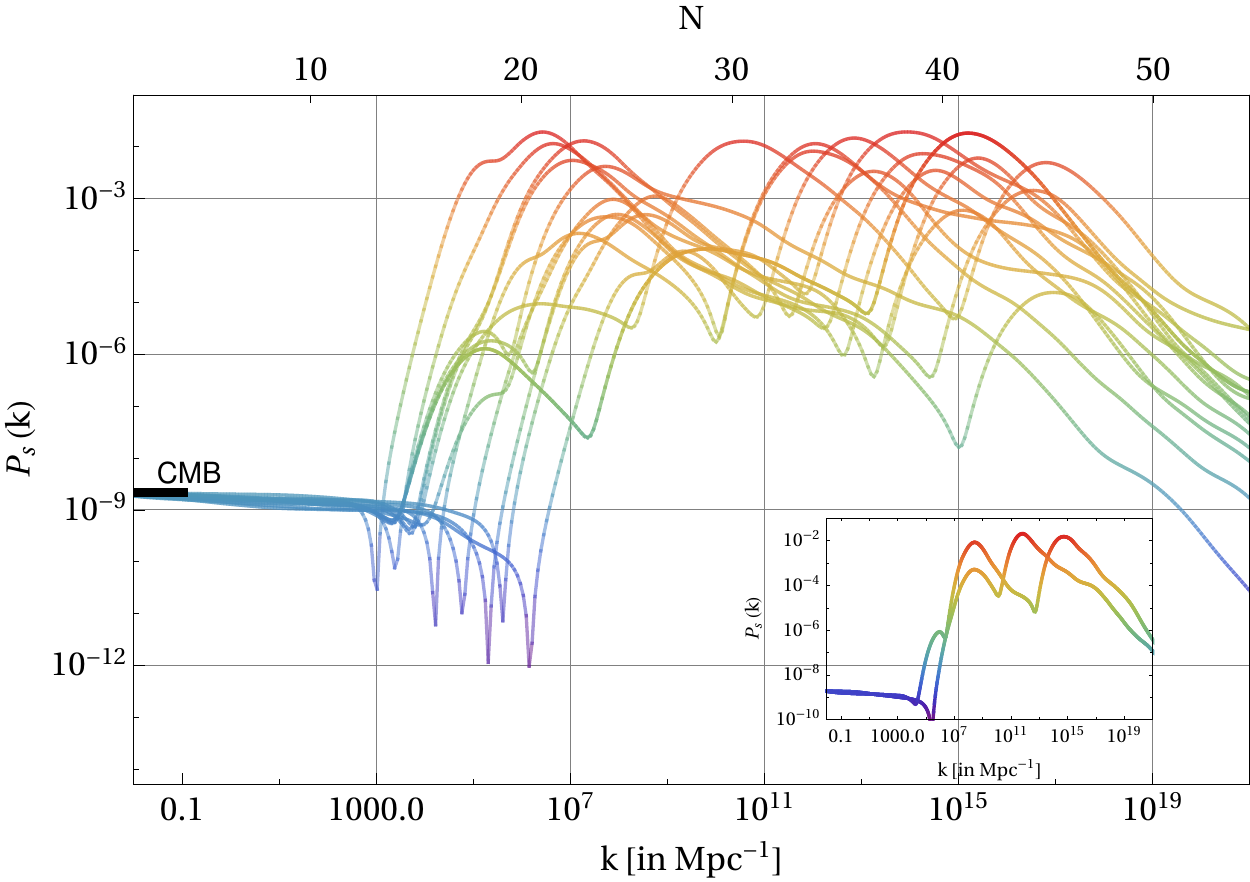}
    \caption{\small \sl This figure represents the predictive 
    outcomes for the power spectrum in 
    the $k$ range $[0.002,10^{21}]\,$Mpc$^{-1}$.  For each 
    prediction, a particular profile is generated starting from 
    the CMB scale. In the inset, we have plotted a couple of 
    profiles to have a better visualization of the generated power spectrum.}
    \label{fig:ps}
\end{figure}

By obtaining predictions for the slow-roll parameters, we can systematically 
assess the power spectrum for each scenario. This analysis enables us to 
ascertain whether these scenarios conform to established cosmological 
patterns or exhibit distinct profiles of the spectrum. 
To derive the power spectrum, we numerically solve the MS equation 
(\ref{u-equation2}) across multiple predictions. The resultant profiles are 
visually presented in fig.\,(\ref{fig:ps}) for 
$k \in [0.002,10^{21}]\,$Mpc$^{-1}$. In that figure, one can discern a 
multitude of profiles, all satisfying the cosmic microwave background (CMB) 
constraint at $k=0.002\,$Mpc$^{-1}$, yet demonstrating diverse evolutionary 
trajectories. Notably, the power spectrum manifests peaks at various $k$ 
values. In numerous instances, multiple peaks (either 2 or 3) are observed, 
a phenomenon conventionally challenging to achieve due to the manual incorporation of 
features in the scalar field potential. Remarkably, in our case, this 
enhancement in multiple predictions occurs naturally, without 
necessitating any specific fine-tuning for the USR period. 
Moreover, the extent of enhancement in the power spectrum varies across 
different predictions, highlighting the flexibility of our framework in 
providing a range of possibilities concerning peak location, amplitude, 
and the number of peaks.

\subsection{Outcomes for PBH Abundance}

\begin{figure}
    \centering
    \includegraphics[height=6cm, width=8.5cm]{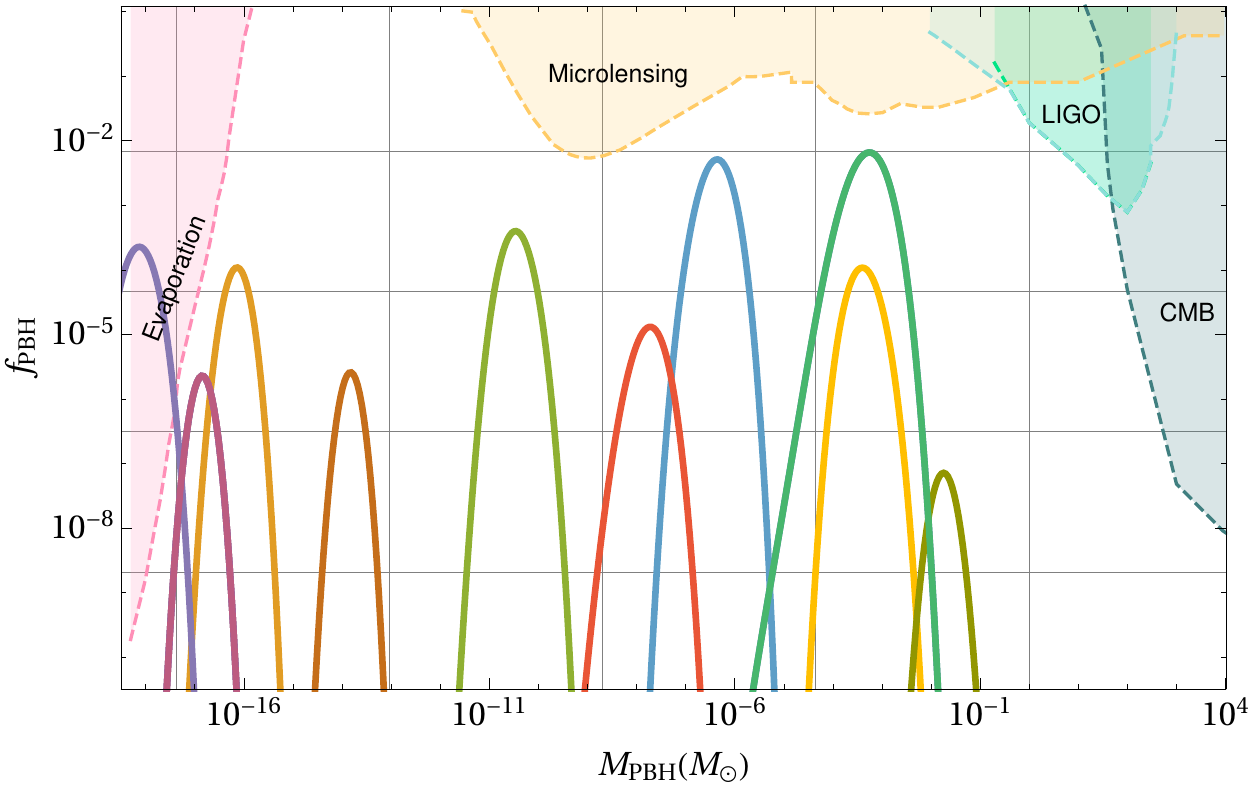}
    \caption{\small \sl This figure illustrates the proportion 
    of dark matter in the form of PBH, denoted as $f_{_{PBH}}$, for various predictions in terms of 
    $M_{_{PBH}}(M_\odot)$. Each vertical contour represents a specific PBH mass with a fraction equal to or less than $f{_{PBH}}$. The shaded regions correspond to areas excluded based on various observations.}
    \label{fig:fpbh}
\end{figure}
\begin{figure}
    \centering
    \includegraphics[height=6cm, width=8.5cm]{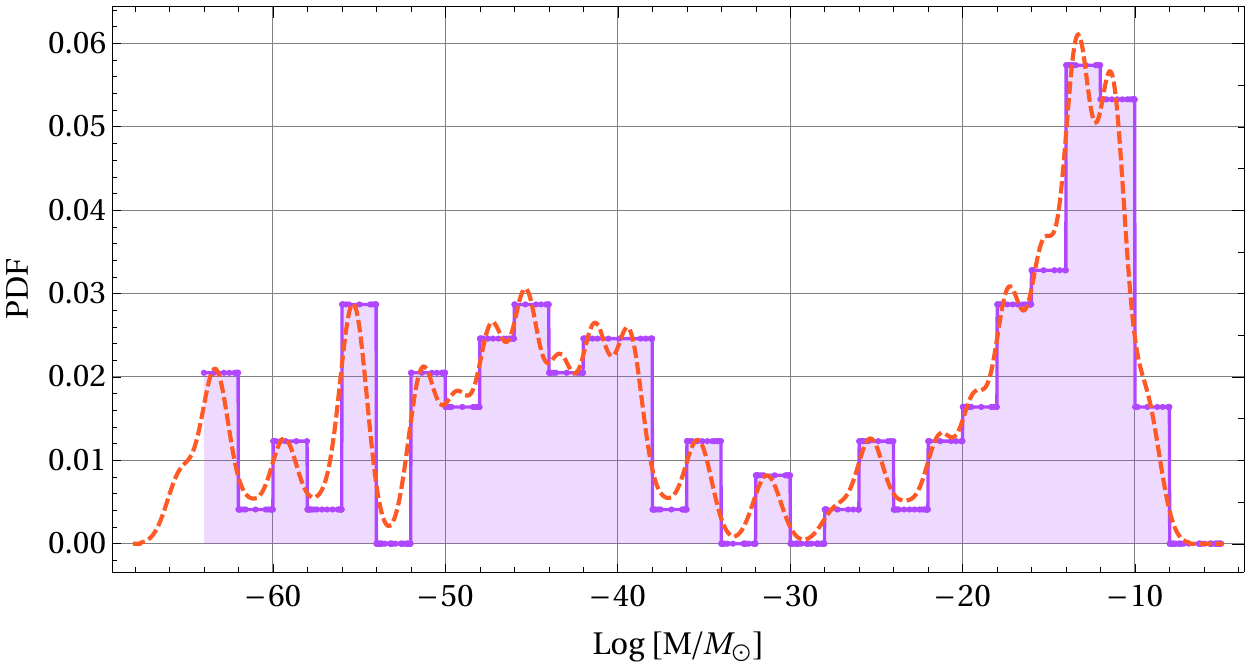}
    \caption{\small \sl This figure illustrates the probability distribution function generated from a substantial number of predictions for obtaining PBHs at a specified mass.}
    \label{fig:dist}
\end{figure}
After obtaining the power spectrum profiles for a set of generated 
predictions, we proceed to derive $f_{_{PBH}}(M_{_{PBH}})$ 
using Eq.\,(\ref{fpbh}) for each specific case. The resulting ensemble 
of predicted contours for $f_{_{PBH}}$ as a function of $M_{_{PBH}}$ in 
the range $[10^{-18},10^{4}]\,M_\odot$ is depicted in fig.\,(\ref{fpbh}). 
In this figure, shaded areas denote excluded regions based on 
various observations, including CMB, GW, micro-lensing, and evaporation. 
Given that certain power spectrum profiles may lead to elevated $f_{_{PBH}}$ 
in specific PBH mass ranges, we have excluded cases that contravene 
observations. Additionally, it is evident from the figure that the peaks 
in the power spectrum generate PBHs over a range of masses rather than a 
specific mass. Furthermore, the abundances fall comfortably within the permissible range of observations.

In fig.\,(\ref{fig:dist}), we have illustrated the probability distribution 
function (PDF) for observing a primordial black hole (PBH) with respect 
to the number of e-foldings ($N$) which corresponding to different masses. 
This distribution is obtained from the $f_{_{PBH}}(M_{_{PBH}})$ distribution 
shown in fig.\,(\ref{fig:fpbh}). Specifically, this figure represents the 
probability of obtaining a PBH of a given mass. A notable observation from 
this figure is that the PDF tends to diminish quite early as it approaches the 
end of inflation. This behavior is a result of the model attempting to 
satisfy the necessary conditions for ending the inflation. The maximum 
likelihood of observing a PBH lies in the range $N \in[22,28]$. This 
might be due to the fact that upto these e-folds, the model strives to 
sustain a sufficient amount of slow-roll period to avoid affecting the CMB 
constraints, and it still has a significant number of e-foldings left to 
smoothly end the inflation. Additionally, a secondary likelihood is observed 
in the power spectrum for $N \in[38,42]$. However, beyond this range, the PDF 
tends to diminish as it approaches $60$ e-foldings, indicating the model is 
trying to ensure a smooth ending to inflation. The observation of a 
non-vanishing PDF in the middle range values of $N$ indicates the 
possibility of a significant enhancement in the power spectrum occurring 
at different locations. This can also be seen in fig.\,(\ref{fig:ps}).

\subsection{Outcomes for SIGW}

\begin{figure}
    \centering
    \includegraphics[height=6cm, width=8.5cm]{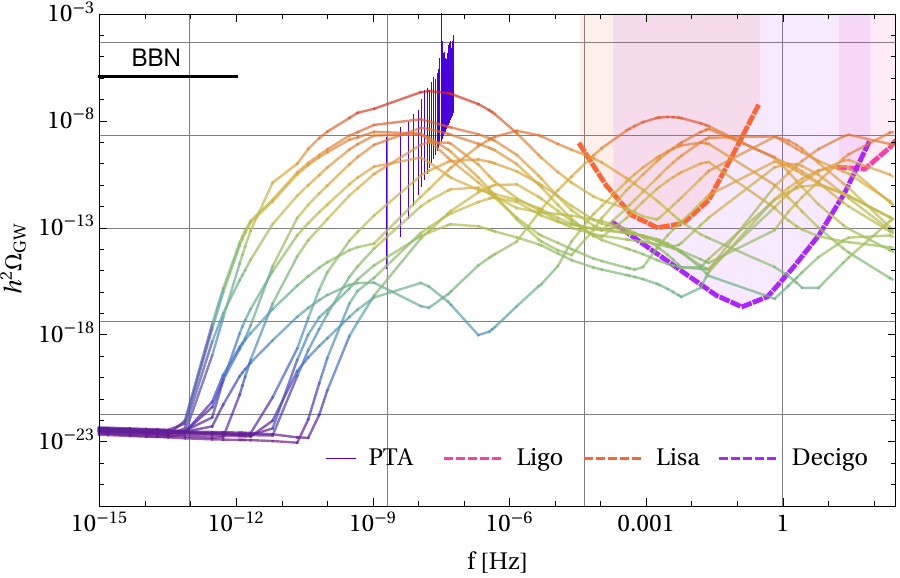}
    \caption{\small \sl This figure showcase the GW spectrum 
    predictions as a function of frequency $f \in[10^{-15},500]
    \,$Hz. The colored regions corresponds to the observational 
    bounds from LIGO, LISA and DECIGO, and the vertical lines at 
    frequency $10^{-9}\,$Hz represent the observational window 
    for the NONAGRAV-15 dataset \cite{NANOGrav:2023,EPTA:2023fyk}.}
    \label{fig:GW}
\end{figure}

To look into the significance of the substantial enhancement 
observed in the scalar power spectrum, we analyze Eq. 
(\ref{GW_spectrum}) to evaluate its potential detection in the 
present-day GW spectrum. The resulting 
outcomes, depicting various predictions for $\Omega_{_{GW}}^{(0)} 
h^2$ across the frequency range $f \in [10^{-15},500]\,$Hz, are 
illustrated in fig.\,(\ref{fig:GW}). In this figure, the 
predictive spectral profiles of GW exhibit distinct behaviors 
from one another but consistently show enhancement in all cases. 
Notably, it is interesting to observe that a majority of these 
spectral profiles fall within the observational window of various 
GW observations such as LIGO, Lisa, Decigo. 
Notably, most of these profiles also meet the observational constraints set by NANOGrav 
\cite{NANOGrav:2023,DeLuca:2020agl}.

It is important to mention that the GW spectra and the current GW observations match well in two different ranges. This is particularly interesting because 
conventional methods typically involve highly unnatural fine-
tuning of parameters to achieve such alignment. In contrast, our 
framework presents a well-behaved scenario that does not rely on 
such assumptions. It allows for the exploration of the 
observational implications of features in the power spectrum 
without the need for precise adjustment 
of parameters.

\section{Conclusion}

In our pursuit to assess the utility of AI algorithm 
in intricate and complex scenarios, such as the generation of 
small-scale primordial perturbations and understanding their 
current cosmological observational evidence, we introduce a novel 
framework in this regard. This framework brings forth intriguing 
possibilities that are typically obscured by the challenges posed by 
severe fine-tuning in such scenarios. On the build framework, 
we train the model using the state-of-the-art Proximal Policy 
Optimization algorithm and the Deep Learning architecture. 
The trained model is then employed to predict scalar power 
spectrum profiles through solving the MS equation. These profiles 
are subsequently utilized for studying various cosmological 
phenomena, including PBH abundance and the generation of SIGW.

The framework comprises a grid that encompasses all possible orders 
of magnitude values for the first slow-roll parameter. The agent's 
transitions between cells determine the second slow-roll parameter 
at a particular instant. 
Through a combination of exploration and exploitation, the agent acquires insights within the environment, determining the optimal transitions at each temporal location, starting from inflation to its end. This learning process enables the agent to fulfil necessary conditions and align with desirable ones. As the agent begins identifying the more rewarding actions 
at specific states, it updates its policy and explores accordingly 
in the subsequent episodes. After completing the training, the most 
recent policy is employed for predictions. However, it is essential 
to note that the predictions remain stochastic, aiming to achieve 
desirable outcomes while primarily focusing on satisfying the 
necessary conditions of inflation.

Once the model is trained, we apply it to generate predictions for the corresponding PBH abundance and the production of secondary-induced GWs. For the form, we utilise the standard Press-Schechter approach. We observe that, due to various peaks in the scalar power spectrum occurring at arbitrary $k$-locations and heights, there is a distribution of PBH abundance with mass. This is in contrast to 
obtaining a single-mass PBH, a common approach in the literature 
achieved by imposing bumps, dips, or inflection points at specific 
locations. Interestingly, most samples in the obtained distribution 
of $f_{_{PBH}}$ with $M_{_{PBH}}$ are away from the excluded regions 
defined by various observational limits. Notably, we not only obtain 
a distribution of mass but also a distribution in the maximum value 
of $f_{_{PBH}}$ for a particular mass. This highlights the intrinsic 
flexibility of the framework to predict outcomes without imposing 
fixed conditions beforehand.

Moreover, we utilize the predicted power-spectrum profiles to assess 
their relevance in light of the current observational constraints on 
the GW spectrum. Specifically, we focus on the scalar induced GW production, 
where the scalar power spectrum undergoes enhancement. Surprisingly, 
the resulting GW spectra for various predictions inherently tend to 
align with the constraints of small frequencies, encompassing NANOGrav, 
as well as those of LIGO, LISA, and DECIGO. This is noteworthy, as in 
many cosmological models, parameters are usually adjusted in a way that 
accommodates only one set of observations at a time. Consequently, our 
framework not only transcends these limitations associated with 
model-based approaches but also identifies profiles that are 
unattainable in those models. In a nutshell, our approach provides a 
range of predictions about inflation, while trying to meet the 
constraints set by various observations.

\section*{Acknowledgement}
DFM thanks the Research Council of Norway for their support and the 
UNINETT Sigma2 -- the National Infrastructure for High Performance Computing and 
Data Storage in Norway. MS is supported by Science
and Engineering Research Board (SERB), DST, Government of India under the Grant Agreement number CRG/2022/004120 (Core Research Grant). MS is
also partially supported by the Ministry of Education
and Science of the Republic of Kazakhstan, Grant No.
0118RK00935, and CAS President’s International Fellowship Initiative (PIFI).

\section*{Appendix-A. Details of RL Framework}\label{app2}

In this appendix, we provide a comprehensive overview of our framework. 
\paragraph{\underline{State-space exploration}:} As already mentioned, our framework comprises a rectangular grid where each cell $C_{i,j}$ contains a specific value of $\epsilon_{_H}$. Alongside defining states, we also need to define a set of actions:
\begin{equation}\label{actions}
    A = \{a_1, a_2, a_3, ... , a_n \} \,,
\end{equation}
where each action $a_i$ determines the extent of change the agent can induce in its state value 
$\epsilon_{_H}$. 
Depending on the choice of the action, the agent can make smaller steps which leads to smaller 
shift in the state value to mimic the slow-roll condition or it can jump over a number of cells 
which leads to a larger change in the state value. This is essential to mimic the bump or dip like 
features in the scalar field potential by enhancing the second slow-roll parameter, $\eta_{_H}$. 

\paragraph{\underline{Reward Structure}:} Initially, when the agent explores the environment, 
it can take any of actions (\ref{actions}) at any state, i.e. it takes any of the actions from a 
random uniform distribution. However, as the agent gathers information regarding 
which cell yields rewards or penalties, it opts to take steps that maximize its overall return 
of a single trajectory using an on-policy method. The agent receives rewards or penalties 
at each timestep based on several factors: ($i$) whether it satisfy $n_s$ and $r_T$ bound imposed 
by Planck 2018, ($ii$) whether it satisfy the slow-roll condition, ($iii$) whether it experiences significant changes in its state value leading to desirable enhancement in 
$\eta_{_H}$, i.e. $>1$, and ($iv$) whether it successfully ends the inflation smoothly. 

Given the infrequent occurrence of condition ($iii$), typically once or twice per rollout, we 
reward the agent more when it significantly reduces $\epsilon_{_H}$ by several orders of 
magnitude. Similarly, we give larger reward for condition ($iv$) to meet the inflation end condition. 
This balances exploration across all required outcomes technically. In particular,
we provide larger reward for these conditions due to their rarity as compared to the ($ii$) condition.

Since slow-roll conditions persist for most timesteps, therefore we give less reward. As these 
conditions must be met for most timesteps, the agent quickly learns actions to which satify slow-roll 
condition.

Actions resulting in conditions ($iii$) and ($iv$) receive rewards almost ten times greater than
those for fulfilling the slow-roll conditions. The reason behind giving reward at each timestep 
is to help the agent to learn faster.

\paragraph{\underline{Model Hyperparameters}:} We have opted for a learning rate of $0.0003$ for our training, discount factor of 
$0.99$, clip range set at $0.2$, entropy coefficient set to $0.01$, batch size of $64$, and ReLU activation function. 

\paragraph{\underline{Exploration vs. Exploitation Strategy}:} During the initial stages of training, 
the agent encourages exploration by utilizing a stochastic policy distribution, allowing the agent to sample a variety of actions to learn about the environment. As training proceeds and the policy improves, the agent naturally reduces exploration and focuses more on exploiting the learned policy to maximize rewards. By opting for a non-zero entropy coefficient, we make sure that the agent continue to explore during training, and avoid overly deterministic policies. This choice is useful to prevent the 
sub-optimal convergence.

\paragraph{\underline{Diversity in predictions}:} As we can see in the predictions shown above 
such as in fig.\,(\ref{fig:ps}), there is a noticeable diversity in the predictions, 
which is expected with the trained on-policy methods. The trained model effectively captures 
the underline probability distribution of different actions for each state within the 
grid which maximizes the overall return. During the prediction phase, the model selects actions based on their probabilities, 
which reflects some level of uncertainty in the decision-making. Therefore, the model does not 
strictly stick to deterministic actions at each state; instead, it leans towards choices according to 
their associated probabilities. It is worth noting that this behavior is not due to incomplete 
training but rather arises from the way the actions are being chosen based on their 
learned probability distributions. Consequently, both the occurrence of the USR period and 
the enhancement in the power spectrum contains some level of stochasticity.

\section*{Appendix-B. Physical interpretation of model's predictions }\label{app1}

\begin{figure}
    \centering
    \includegraphics[height=6cm, width=8.5cm]{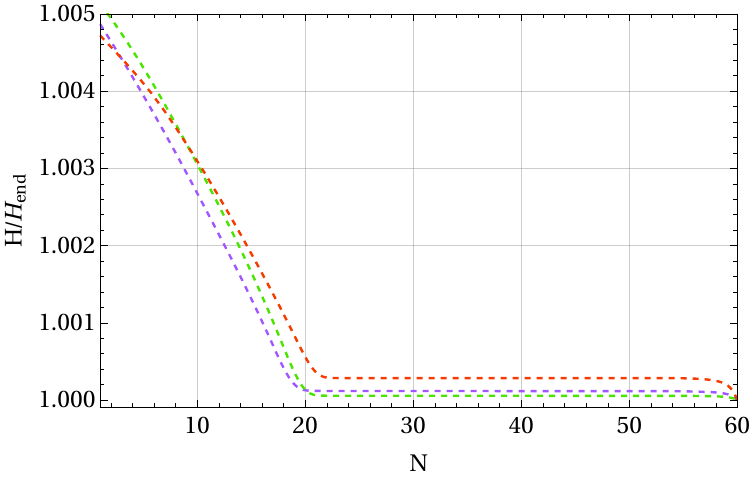}
    \caption{Evolution of $H/H_\text{end}$ with number of e-foldings $N \in [0,60]$ for some of the 
    model's predictions. }
    \label{fig:HoHend}
\end{figure}

In order to see the compatibility of the obtained predictive profiles of primordial power spectrum 
with the actual inflationary phenomena (with an ultra slow-roll period) we need to see how 
does the Hubble parameter is constructed out. Given the functional profiles of $\epsilon_{_H}$, one can directly constructed out the Hubble parameter evolution \cite{Ozsoy:2023ryl} 
\begin{equation}
    H(N) = H_\text{end} \exp\left[- \int_{60} ^N \epsilon(N') dN' \right] \,.
\end{equation}
In the figure (\ref{fig:HoHend}) we demonstrated the functional profiles of $H(N)/H_\text{end}$ 
for some of the predictions on $\epsilon_{_H}$. Here, one can see that the Hubble parameter decreases 
in the beginning of the inflation and quickly approaches to $H_\text{end}$, 
where its variation is minuscule. This outcome is essential to realize the slow-roll regime. 

To compare our results with the scalar field's inflationary scenario, we can use 
the standard equations:
\begin{eqnarray}
    H^2(t) &=& \frac{1}{3} \left[\frac{\dot{\phi}^2}{2} + V(\phi) \right] \,, \\
    \dot{H}(t) &=& -\frac{1}{2} \dot{\phi}^2  \,. 
\end{eqnarray}
Given that $\dot{H}(t)$ is very small from fig.\,(\ref{fig:HoHend}), it is evident that the field's 
kinetic energy is insignificant in comparison to its potential energy. This validates the physical credibility of our results by mimicking the standard inflationary scenario. 

The above fig.\,(\ref{fig:HoHend}) indicates that to mimic the slow-roll or USR regime
one does not need an explicitly form of with scalar field potential. Instead, one can 
exploit the general idea of satisfying the necessary inflationary conditions to analyze the 
features of the inflation.

\end{document}